\documentclass[cameraready]{Interspeech}
\title{Read What You Hear: Reference-Free Hypotheses Evaluation with Acoustic Discrepancy}
\author{Zhihan}{Li}
\author{Hankun}{Wang}
\author{Yiwei}{Guo}
\author{Bohan}{Li}
\author{Xie}{Chen}
\author[correspondingauthor]{Kai}{Yu}
\address{
X-LANCE Lab, School of Computer Science, Shanghai Jiao Tong University, China \\
MoE Key Lab of Artificial Intelligence, Jiangsu Key Lab of Language Computing, China
}

\email{\{lizhihan957, kai.yu\}@sjtu.edu.cn}


\usepackage{comment}
\usepackage{multirow}
\usepackage{graphicx}
\usepackage{subcaption}
\usepackage{makecell}

\begin{document}

\maketitle

\begin{abstract}
Automatic speech recognition systems commonly rely on reference transcriptions for evaluation, while reference-free approaches often depend on internal confidence estimation or auxiliary language models. We propose READ (Reference-free Hypothesis Evaluation with Acoustic Discrepancy), a novel metric that evaluates ASR hypotheses directly from the speech signal. READ emphasizes the acoustic grounding of hypotheses. It uses a pretrained auto-regressive TTS model to compute the conditional likelihood of speech tokens given a text hypothesis, to measure fine-grained acoustic discrepancy between speech and text. Without additional training, READ can be applied for hypothesis refinement. Experiments show that READ correlates with specific recognition errors and improves ASR outputs, achieving up to 20\% relative error rate reduction, with particularly strong gains under noisy conditions.
% Automatic speech recognition systems commonly rely on reference transcriptions for evaluation and optimization, limiting their applicability in unlabeled or low-resource scenarios, while reference-free approaches often depend on internal confidence estimation or auxiliary language models. In this work, we propose a reference-free method for ASR hypothesis evaluation and refinement based on TTS-based speech-text consistency. We leverage the loss of a text-to-speech model as an external and model-agnostic indicator of speech-text consistency.
% Experiments on single-system N-best rescoring show that ASR error rate can be relatively reduced by 2\%-20\% through TTS-based evaluation. In multi-system combination, the proposed method can generate hypotheses better than any individual system or ROVER baseline. Our method yields additional improvements on noisy datasets, demonstrating the robustness advantage of the proposed method.
\end{abstract}

\section{Introduction}

% Automatic Speech Recognition (ASR) has achieved remarkable progress with the advent of large-scale pre-training and end-to-end architectures. However, \textbf{hypothesis evaluation} of ASR systems remains a non-trivial challenge, particularly in real-world scenarios where ground-truth transcripts are often unavailable. Traditional evaluation relies heavily on reference-based metrics such as Word Error Rate (WER). While WER provides a definitive measure of accuracy, its requirement for human-labeled references limits its utility large-scale unsupervised scenarios and iterative system self-improvement.

Automatic Speech Recognition (ASR) has achieved remarkable progress with the advent of large-scale pre-training and end-to-end architectures. However, \textbf{hypothesis evaluation} of ASR systems remains a non-trivial challenge, particularly in real-world scenarios where ground-truth transcripts are often unavailable. Traditional evaluation relies heavily on reference-based metrics such as Word Error Rate (WER). While WER provides a definitive measure of accuracy, its requirement for human-labeled references limits its utility in large-scale unsupervised scenarios and iterative system self-improvement.

% To address the absence of references, various reference-free evaluation methods have been proposed. Internal confidence scores~\cite{wessel2001confidence,jiang2005confidence} derived from ASR decoders offer a computationally efficient proxy, but often suffer from poor calibration and overconfidence~\cite{zhang2001word,guo2017calibration}. External evaluation methods such as Language Model (LM) rescoring~\cite{chelba2012large,mikolov2010recurrent}, incorporate linguistic priors to assess the ``fluency" of hypotheses. However, text-only approaches inherently overlook the speech signal. Quality Estimation (QE) models like eWER~\cite{Ali2018eWER} have attempted to predict absolute error rates, yet they typically require supervised training on error-labeled datasets and lack the granularity to pinpoint the exact error locations.

To address the absence of references, various reference-free evaluation methods have been proposed. Internal confidence scores~\cite{wessel2001confidence,jiang2005confidence} derived from ASR decoders offer a computationally efficient proxy, but often suffer from poor calibration and overconfidence~\cite{zhang2001word,guo2017calibration}. External evaluation methods such as Language Model (LM) rescoring~\cite{chelba2012large,mikolov2010recurrent}, incorporate linguistic priors to assess the ``fluency" of hypotheses. However, text-only approaches inherently overlook the speech signal. Quality Estimation (QE) models like eWER~\cite{Ali2018eWER} have attempted to predict absolute error rates, yet they typically require supervised training on error-labeled datasets and lack the granularity to pinpoint the exact error locations.

% A primary application of evaluation methods is \textbf{hypothesis refinement}, adjusting ASR outputs to achieve lower error rates. Classical refinement strategies, such as $N$-best reranking and system combination (e.g., ROVER~\cite{rover}), often utilize confidence scores or rescoring weights. 
% The emergence of Large Language Models (LLMs) has introduced a generative paradigm, where the model directly produces final results without explicit evaluation~\cite{ma2023can,chen2023hyporadise,tur2024progres}. While generative approaches demonstrate impressive performance, we argue that refinement cannot replace evaluation. Beyond pursuing a better output, we need evaluation metrics that provide interpretable and diagnostic insights to guide the optimization of existing systems.

A primary application of evaluation methods is \textbf{hypothesis refinement}, adjusting ASR outputs to achieve lower error rates. Classical refinement strategies, such as $N$-best reranking and system combination (e.g., ROVER~\cite{rover}), often utilize confidence scores or rescoring weights. 
The emergence of Large Language Models (LLMs) has introduced a generative paradigm, directly producing improved hypotheses without explicit evaluation~\cite{ma2023can,chen2023hyporadise,tur2024progres}. While generative approaches demonstrate impressive performance, they do not eliminate the need for evaluation. We still need metrics that provide interpretable and diagnostic insights to guide the optimization of existing systems.

% We believe that a system should be cognizant of its own actions, so do the evaluation methods.~\cite{xu2024rejection} emphasizes the significance of model reliability and proposed specific metrics. Inspired by these insights, we reconsider what reliability means in ASR. Recalling the early formulation of ASR, Bayes' theorem decomposes the posterior $P(\text{text} \mid \text{speech})$ into an acoustic model $P(\text{speech} \mid \text{text})$ and a language model $P(\text{text})$. While LM rescoring reinforces \textbf{linguistic plausibility}, a corresponding mechanism for \textbf{acoustic grounding} is notably absent. Even when speech signals are incorporated in some works~\cite{ali2020word,park2025fast,waheed2025robust,hu2024large,hu2024listen}, they are processed within E2E black boxes, precluding a clear verification of acoustic versus linguistic contributions.

We believe that a system should be cognizant of its own actions, so do the evaluation methods.~\cite{xu2024rejection} emphasizes the significance of model reliability and proposed specific metrics. Inspired by these insights, we reconsider what reliability means in ASR. Recalling the early formulation of ASR, Bayes' theorem decomposes the posterior $P(\text{text} \mid \text{speech})$ into an acoustic model $P(\text{speech} \mid \text{text})$ and a language model $P(\text{text})$. While LM rescoring reinforces \textbf{linguistic plausibility}, a corresponding mechanism for \textbf{acoustic grounding} is notably absent. Though speech signals are incorporated in some works~\cite{ali2020word,park2025fast,waheed2025robust,hu2024large,hu2024listen}, acoustic and linguistic factors are processed within E2E black boxes, hindering clear verification.

\begin{figure}[t]
  \centering
  \includegraphics[width=\linewidth]{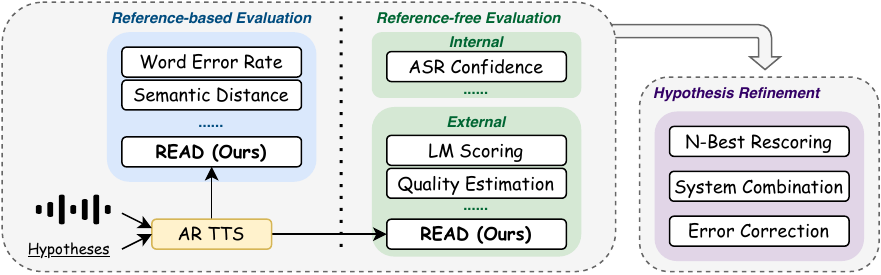}
  \vspace{-0.2in}
  \caption{We propose READ for hypothesis evaluation.}
  \label{fig:summary}
  \vspace{-0.2in}
\end{figure}

In this paper, we propose \textbf{READ} (\textbf{R}eference-free Hypothesis \textbf{E}valuation with \textbf{A}coustic \textbf{D}iscrepancy), a novel unsupervised metric that restores the balance by re-emphasizing the role of the acoustic model. Our motivation stems from the principle of \textit{analysis-by-synthesis}~\cite{halle2003speech}: if a hypothesis accurately represents the speech, we should be able to resynthesize the speech with high likelihood. By utilizing a pre-trained discrete auto-regressive TTS model (e.g., CosyVoice2~\cite{du2024cosyvoice2}) in a teacher-forcing mode, we compute the conditional likelihood of discrete speech tokens given a text hypothesis.
The resulting negative log-likelihood (NLL) sequence serves as a fine-grained ``acoustic discrepancy'' map, where spikes in loss directly correspond to misaligned or erroneous segments in the ASR output. READ places a strong emphasis on \textbf{acoustic grounding}, ensuring that the hypothesis is faithful to the speech signal. Our proposed READ metric offers the following advantages:

\begin{itemize}
\item \textbf{Reference-free evaluation.} 
READ can be used without relying on ground truth. By evaluating the discrepancy between a hypothesis and the speech signal, READ is able to assess relative quality among arbitrary hypotheses.

\item \textbf{Locality modeling.} 
READ produces metrics aligned with the speech frame sequence and concentrates acoustic discrepancies within their corresponding temporal regions. This enables precise localization of specific errors.

\item \textbf{Model-agnostic and training-free.} 
READ requires no additional training on specific ASR systems or target datasets. The evaluation relies entirely on the intrinsic knowledge in an off-the-shelf TTS model.

\item \textbf{Robustness against noisy speech.} Benefiting from the stable acoustic modeling from TTS systems, READ-based evaluation demonstrates stronger performance when confronted with challenging acoustic conditions, such as noisy speech.
\end{itemize}

\section{Background}

\subsection{ASR Hypothesis Evaluation}
ASR Hypothesis Evaluation concerns the problem of assessing how well the hypothesis explains the original speech signal.

\textbf{Reference-based Evaluation.} Reference-based methods directly compare the generated hypothesis with a given ground-truth transcript. Among these, the edit-distance-based Word Error Rate (WER) has become a foundational metric. Alternative measures such as Semantic Distance~\cite{kim21e_interspeech} evaluates the text discrepancy in semantic space rather than lexical space.

% Reference-based methods are well suited for benchmarking existing ASR systems, as well as for supervised fine-tuning and reinforcement learning scenarios. However, their applicability is constrained when ground-truth transcripts cannot be massively obtained, limiting their scalability in low-resource or unlabeled environments.

\textbf{Reference-free Internal Evaluation.}
Focusing on confidence estimation derived from the internal signals of ASR systems, hypothesis reliability is inferred from posterior probabilities, hidden representations, etc. Such methods are computationally efficient and naturally integrated into the decoding process. However, internal confidence relies entirely on the model, it often suffers from calibration issues and overconfidence~\cite{wessel2001confidence,jiang2005confidence,futami2021asr,oneactua2021evaluation,zhang2001word,guo2017calibration}.

\textbf{Reference-free External Evaluation.}
External evaluation introduces auxiliary knowledge sources to assess hypothesis quality. A common strategy is Language Model (LM) rescoring, leveraging linguistic prior probabilities to evaluate the hypotheses. Quality Estimation (QE) methods~\cite{Ali2018eWER,ali2020word,park2025fast,waheed2025robust} train proxy models to approximate metrics like WER without grounding truth.

% Although QE methods attempt to provide absolute and system-agnostic quality indicators, their performance is highly dependent on the training data distribution. Moreover, they typically offer limited fine-grained error localization, restricting their interpretability and diagnostic utility.

\textbf{Hypothesis Refinement.}
Hypothesis refinement refers to methods that directly refine the final output. Classical refinement approaches include reranking based on LM rescoring or ASR confidence, as well as system combination methods such as ROVER~\cite{rover}. Recently, the emergence of LLMs has enabled generative revising paradigms, in which explicit evaluation is bypassed and LLM directly produces refined transcriptions.

\subsection{Auto-Regressive Text-to-Speech Systems Based on Discrete Tokens}

Many works~\cite{wang2023neural,du2024cosyvoice2,du2025cosyvoice,deng2025indextts} model TTS by auto-regressively predicting discrete speech tokens. They typically rely on a causal Transformer to model the distribution of the next token. As a result, such models can naturally provide the conditional likelihood of token sequences for evaluation. Moreover, their auto-regressive component is often based on semantic tokens, focusing primarily on the linguistic content of the speech signal. This makes them more task-relevant for ASR evaluation.

% TTS research has witnessed the emergence of a new paradigm based on discrete speech representations. Speech signals are first converted into discrete tokens before sequence modeling.

% \textbf{Semantic Speech Tokens.}
% Speech tokens can be roughly categorized into two types[]. In contrast to acoustic tokens, semantic tokens are obtained from self-supervised learning (SSL) models or tokenizers trained with non-reconstruction objectives such as ASR. These tokens focus primarily on linguistic content while discarding most paralinguistic information. Representing speech at the semantic-token level could be more task-relevant for ASR evaluation.

% \textbf{Auto-Regressive Speech Modeling.} Auto-regressive TTS models are typically trained with a causal Transformer decoder under the next-token prediction (NTP) objective, modeling the conditional probability distribution of the next token. This formulation naturally yields the likelihood of arbitrary token segments, enabling the computation of conditional probabilities for any subsequence.

% Representative semantic-token-based auto-regressive TTS models include the CosyVoice series, Chatterbox, and the IndexTTS series. Our experiments are primarily conducted on CosyVoice2, which adopts S3Tokenizer—a supervised tokenizer trained with ASR-oriented objectives.

\section{READ: Reference-Free Hypotheses Evaluation with Acoustic Discrepancy}

\subsection{Deriving Acoustic Discrepancy from AR TTS Systems}
Let $X = (X_1,\dots,X_N)$ denote a sequence of text tokens and $Y = (Y_1,\dots,Y_T)$ denote a sequence of speech tokens. A trained auto-regressive TTS system parameterized by $\theta$ models the sequence causally, defining a conditional probability distribution $P_\theta(Y_t \mid X, Y_{<t})$ at each time step $t$, where $Y_{<t}=(Y_1,\dots,Y_{t-1})$. This enables likelihood evaluation for any speech-text pair $(x,y)$ regardless of whether it is matched. We define the frame-level conditional likelihood at step $t$ as
\begin{align}
\ell_t(x,y)
:=
P_\theta(Y_t = y_t 
\mid X=x, Y_{<t}=y_{<t}),
\end{align}
which measures the consistency of the observed speech token $y_t$ with the conditioning text and preceding acoustic context.

Due to maximum likelihood training, the TTS model explicitly maximizes $\ell_t$ at each time step for matched speech-text pairs. Therefore, speech-text pairs that are consistent with the actual distribution are expected to yield higher values of $\ell_t$, whereas mismatched pairs typically result in lower $\ell_t$.

% this is an occasional name to change
We define \textbf{READ} (\textbf{R}eference-Free Hypotheses \textbf{E}valuation with \textbf{A}coustic \textbf{D}iscrepancy) as
\begin{align}
\mathrm{READ}_t(x,y) = -\log \ell_t,
\mathrm{READ}(x,y) =\sum_{t=1}^T \mathrm {READ}_t.
\end{align}
This defines a sequence of discrepancy scores aligned with the speech sequence. Notice that $\mathrm {READ}(x,y) = \sum_{t=1}^{T} -\log \ell_t = -\log P_\theta(y \mid x)$, which corresponds to the negative conditional log-likelihood of the entire speech, and thus serves as a global measure of speech-text discrepancy.

Importantly, the $\mathrm{READ}_t$ sequence exhibits \textbf{locality}. Discrepancies occurring around frame $y_t$ will primarily affect $\mathrm{READ}_t$ and its neighboring positions. As a result, local speech-text mismatches tend to manifest as concentrated increases in $\mathrm{READ}_t$ within the corresponding temporal region. This locality enables $\mathrm{READ}_t$ to provide finer-grained, frame-level indications of speech-text discrepancy.

For fine-grained evaluation, it is desirable to transfer $\mathrm{READ}_t$ defined along the speech sequence $y$ back to the text sequence $x$ using the monotonic alignment between $y$ and $x$.
% In practice, text tokens and speech tokens typically exhibit a monotonic one-to-many relationship, which can be modeled through a monotonic alignment. 
Multiple alignment models~\cite{moreno1998recursive,povey2011kaldi,shi2026qwen3} can provide such mappings. However, these alignments originate from systems that are different from the TTS model for measuring $\mathrm{READ}_t$, potentially introducing cross-model inconsistencies.
To maintain internal consistency, we instead extract alignment directly from the same auto-regressive TTS model that derives $\mathrm{READ}_t$. 

Following prior work~\cite{wang2024attention}, we discover and utilize the speech-text alignment information from the attention heads in the decoder-only Transformer of a TTS model.
Let the concatenated input sequence have length $T+N$, consisting of $N$ text tokens followed by $T$ speech tokens under causal masking. The full self-attention map is therefore a $(T+N)\times(T+N)$ matrix. From this matrix, we extract the $T\times N$ submatrix corresponding to attention weights from speech positions to text positions, denoted as $A \in \mathbb{R}^{T \times N}$,
where $A_{t,n}$ represents the attention weight from speech frame $t$ to text token $n$.
We formulate alignment as a monotonic mapping $ \pi : \{1,\dots,T\} \rightarrow \{1,\dots,N\}$,
% \begin{align}
% \pi : \{1,\dots,T\} \rightarrow \{1,\dots,N\},
% \end{align}
with monotonicity constraint $\pi(t+1) \ge \pi(t)$ and boundary conditions $\pi(1)=1, \pi(T)=N$. The optimal alignment is obtained via dynamic programming~\cite{wang2024attention} as
\begin{align}
\pi^*
=
\arg\max_{\pi \in \mathcal{M}}
\sum_{t=1}^{T} A_{t,\pi(t)},
\end{align}
where $\mathcal{M}$ denotes the set of all possible mappings satisfying the above constraint.

Given the optimal mapping $\pi^*$, we aggregate discrepancy from speech-token-level to any text segment:
\begin{align}\label{eq:align}
\mathrm{READ}^{\text{text}}_{[n_1,n_2]}
=
\sum_{t:\,n_1\leq \pi^*(t)\leq n_2}
\mathrm{READ}_t.
\end{align}

This yields discrepancy scores for any text segments, enabling fine-grained identification of positions that are poorly supported by the speech signal.

\subsection{ASR Hypothesis Refinement with READ}
\label{sec:refinement-method}
\subsubsection{Sentence-Level Rescoring}
The proposed READ metric can first be used for sentence-level hypothesis rescoring.
Given an $N$-best list or multiple candidate hypotheses from multiple ASR systems, we compute the total acoustic discrepancy for each candidate text $x^{(i)}$ conditioned on the same speech sequence $y$:
\begin{align}
\mathrm{Score}\left(x^{(i)} \mid y\right)=-\mathrm {READ}\left(x^{(i)},y\right).
\end{align}
As previously mentioned, this corresponds to the log-likelihood of the entire speech under different hypotheses. Hypotheses with smaller total acoustic discrepancy are rescored higher. 

\subsubsection{Segment-Level Combination}
\label{subsubsec:segment_level_combination}
\begin{figure}[t]
  \centering
  \includegraphics[width=\linewidth]{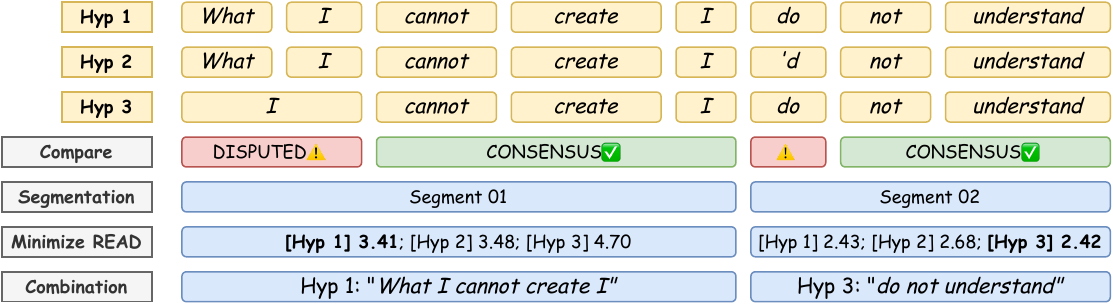}
  \caption{Segment-level combination with READ Evaluation.}
  \label{fig:segment_level_combination}
  \vspace{-0.2in}
\end{figure}

Since READ exhibits locality, it enables finer-grained system combination.
To achieve this, we perform segmentation along the speech signal, as shown in Figure~\ref{fig:segment_level_combination}. We first identify intervals where different hypotheses dispute and those where they reach consensus. Since READ is temporally aligned, an interval is considered consensus only when both the content and the time span are identical across hypotheses.

We assume that the influence of a disputed interval persists at most until the beginning of the next disputed interval, covering itself together with a subsequent consensus interval. Segmentation is performed according to this strategy and Eq.\eqref{eq:align}. 
We select the hypothesis that yields minimal READ value for each segment.

Although this strategy relies on a locality assumption and adopts a greedy selection scheme, experimental results show that more than 98\% of the merged outputs achieve simultaneous READ reduction within each segment.

\subsubsection{Integrating READ with ROVER}
Moreover, we show that READ can also corporate with other system combination methods, like ROVER~\cite{rover}.
Since READ sequences are not directly comparable across segments of unequal length, combination cannot be performed at a granularity finer than the intervals in Section~\ref{subsubsec:segment_level_combination}. To address this, we adopt a simple strategy: we treat the segment-level combination as a new candidate and feed it into ROVER together with the original candidates. The system retains ROVER’s token-level merging capability, while the new candidate injects a segment-level bias.

% \begin{figure}[t]
%   \centering
%   \includegraphics[width=\linewidth]{pics/Integrating_ROVER.png}
%   \caption{Integrating READ into ROVER.}
%   \label{fig:integrating_rover}
% \end{figure}

% Since we do not introduce any confidence learning or quality estimation training, ROVER itself adopts equal-weight voting. By tuning the weights of segment-level combination, we can introduce varying degrees of acoustic bias into ROVER.

\section{Experiments}

\subsection{Experimental Setup}
We conduct experiments mainly on CosyVoice2~\cite{du2024cosyvoice2}, a discrete auto-regressive TTS system. We adopt the official checkpoints, which have not been subjected to any additional training on the involved datasets. This setup is to make sure that the evaluation capability originates purely from the TTS model itself.

For candidate ASR systems, we select powerful models with different architectures, including Whisper~\cite{whisper} (medium and large-v3), NVIDIA NeMo~\cite{nemofastconformer}, and Qwen2.5-Omni~\cite{xu2025qwen25omnitechnicalreport}. We adopt Whisper-large-v3 in $N$-best generation. Following~\cite{chen2023hyporadise}, we perform beam search decoding with a beam size of 60, then remove duplicates caused by case, punctuation and timestamps. Finally we keep the top-5 unique candidates.

For test-sets, we use the LibriSpeech~\cite{librispeech} test-clean and test-other splits as representatives of clean speech data. SPGISpeech~\cite{spgispeech}, Switchboard~\cite{godfrey1992switchboard}, TEDLIUM3~\cite{tedlium3}, and VCTK-noisy~\cite{valentini2017vctknoisy} are employed to evaluate performance under diverse acoustic scenarios. We further consider the challenging code-switching setting, and include the ASRU2019~\cite{shi2020asru} and TALCS~\cite{talcs} Mandarin–English code-switching datasets.

To simulate more challenging acoustic conditions, we construct noise-augmented versions of the aforementioned datasets. Random noise from the WHAM!~\cite{wham} test set is added at SNR levels of 0, 10, and 20dB.
\subsection{Effectiveness of READ Evaluation}

\subsubsection{READ for Reference-Based Evaluation}

\begin{figure}
    \centering
    \includegraphics[width=\linewidth]{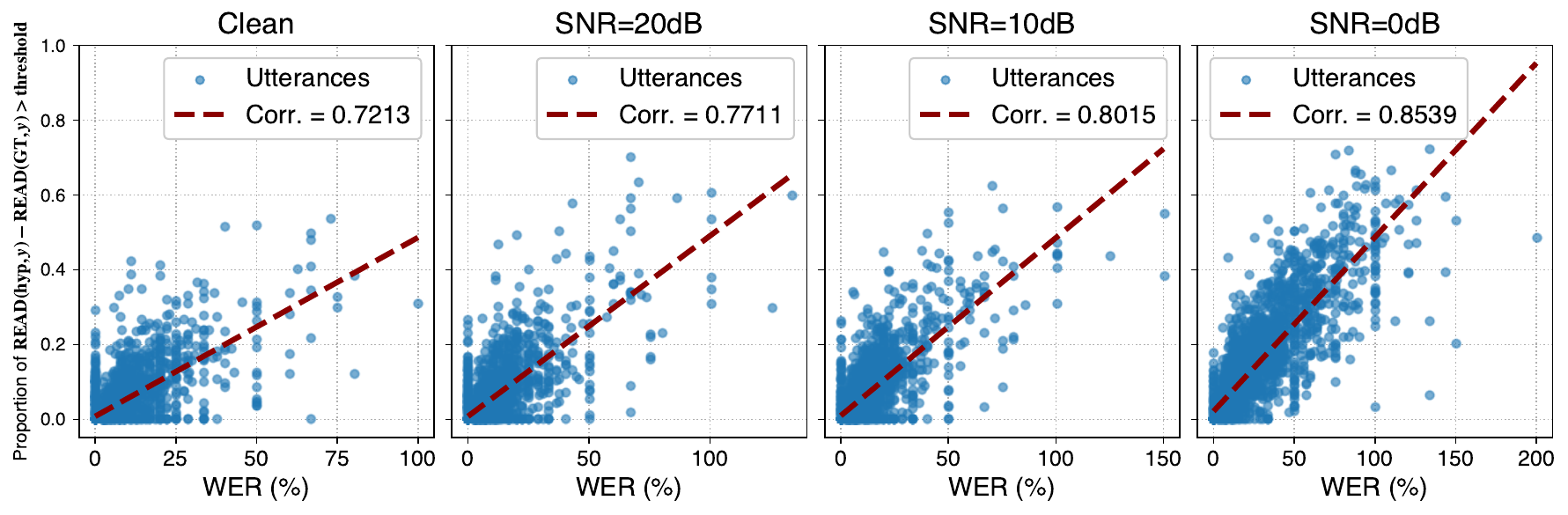}
    \vspace{-0.2in}
    % \caption{Correlation between READ difference and WER. The y-axis represents the proportion of duration where the READ value of hypothesis is higher than that of GT by a threshold.}
    \caption{Correlation between READ difference and WER. The y-axis shows the proportion of duration where the difference exceeds an empirically set threshold (0.3).}
    \label{fig:correlation_read_wer}
    \vspace{-0.2in}
\end{figure}

% We first demonstrate the effectiveness of READ in reference-based evaluation. For each hypothesis-reference pair, we compute their difference in READ sequence. A higher READ value for the hypothesis indicates more severe errors.

% We use the proportion of duration where the READ difference exceeds a certain threshold as an error-rate-like metric. Figure~\ref{fig:correlation_read_wer} shows the relationship between this metric and WER. Under different noise conditions, it exhibits a certain degree of correlation, while still differing from WER due to the gap between acoustic modeling and lexical errors. As noise level increases, recognition errors are more likely to stem from failures in acoustic modeling rather than language modeling. 
% Consequently, the correlation between READ and WER becomes stronger, reflecting READ’s bias to acoustic modeling.

We first demonstrate the effectiveness of READ in reference-based evaluation. For each hypothesis-reference pair, we compute the READ difference sequence, where higher values indicate more severe errors.

We define an error-rate-like metric as the proportion of duration where the READ difference exceeds a threshold (set to 0.3 empirically). As shown in Figure~\ref{fig:correlation_read_wer}, this metric correlates with WER across noise conditions, while remaining distinct due to the gap between acoustic and lexical errors. As noise increases, errors are more dominated by acoustic failures, leading to stronger correlation and reflecting READ’s bias toward acoustic modeling.

\subsubsection{Case Study: On the Locality of READ Evaluation}

We illustrate the reference-free evaluation capability of READ and its locality through a case study, considering Hyp~1 from Qwen2.5-Omni and Hyp~2 from Whisper medium on an utterance from SPGISpeech.

\begin{figure}[t]
  \centering
  \includegraphics[width=\linewidth]{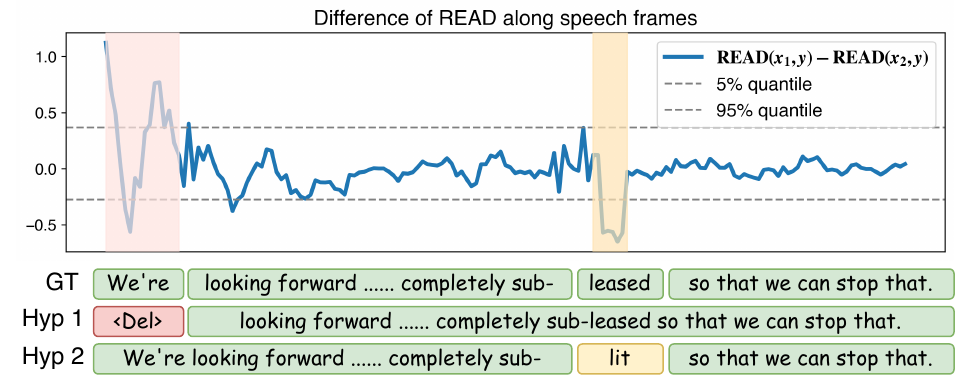}
  \vspace{-0.25in}
  \caption{Case study of READ's locality in reference-free evaluation. When two hypotheses differ, their READ values exhibit clear peaks of difference in the corresponding temporal range.}
  \label{fig:locality}
  \vspace{-0.2in}
\end{figure}

Figure~\ref{fig:locality} presents the difference between the $\mathrm{READ}_t$ sequences of two hypotheses. 
Positive spikes clearly correspond to time regions where Hyp~2 is better supported by the speech signal, while negative ones indicate that Hyp~1 is better. 
With the internal alignment extracted from the TTS model, these peaks can be mapped back to text segments, making it straightforward to locate the words responsible for the mismatch.

This locality endows READ with a fine-grained advantage: 
it can precisely identify where recognition performance deteriorates, 
thus enabling targeted optimization. 
But the capability of READ to distinguish among substitution, deletion, and insertion errors remains to be further explored.

\subsection{Performance of Hypothesis Refinement with READ}
\label{subsec:revising}
% We first conduct reference-free evaluation at dataset-level, to determine a base system with lowest averaging READ. Because READ focuses solely on acoustic grounding, we introduce a small bias (0.95×) on READ toward the base system in subsequent experiments, to maintain a basic level of language modeling, allowing READ-based evaluation to intervene only when acoustic reliability is questionable.

We first perform dataset-level reference-free evaluation to select the base system with the lowest average READ. As READ captures only acoustic grounding, we slightly bias the base system by scaling its READ sequence by 0.95. This retains basic language modeling, allowing READ-based evaluation to act only when acoustic reliability is uncertain.

Then, we use READ for hypothesis refinement in different granularities, illustrated by Section~\ref{sec:refinement-method}.
All results are reported with WER(\%) for English test sets and Mixed Error Rate (MER)(\%) for code-switching test sets.

\subsubsection{Sentence-Level Rescoring}

\begin{table}[h]
\vspace{-5pt}
  \caption{N-Best rescoring performances. ``Oracle" refers to collecting the hypotheses with lowest error rate for each utterance, which is the topline for rescoring.}
  \vspace{-0.1in}
  \setlength{\tabcolsep}{3pt}
  \label{tab:sentence_level}
  \centering
  \scriptsize
  \resizebox{\linewidth}{!}{
  \begin{tabular}{cccccccc} % 改为 8 个 c
    \toprule
    \multirow{2}{*}{\textbf{Dataset}} 
        & \multicolumn{5}{c}{\textbf{Hypothesis}} 
        & \multirow{2.5}{*}{\makecell[c]{\textbf{Rescoring}\\\textbf{w/ READ}}} 
        & \multirow{2}{*}{\textbf{Oracle}} \\
    \cmidrule(lr){2-6}
    & 1st & 2nd & 3rd & 4th & 5th\\
    \midrule
    LS-clean & \textbf{2.06} & 2.74 & 3.10 & 3.57 & 4.12 & \textbf{1.91~(-7.28\%)} & 1.15\\
    LS-other & \textbf{3.66} & 4.71 & 5.37 & 5.63 & 6.06 & \textbf{3.48~(-4.92\%)} & 2.15 \\
    VCTK-noisy & \textbf{7.41} & 8.76 & 9.21 & 10.85 & 12.18 & \textbf{7.19~(-2.97\%)} & 6.14 \\
    ASRU-test & \textbf{9.96} & 11.97 & 12.64 & 13.39 & 14.42 & \textbf{9.67~(-2.91\%)} & 6.20 \\ 
    TALCS-test & 18.94 & \textbf{18.56} & 18.57 & 18.93 & 19.69 & \textbf{14.98~(-20.91\%)} & 13.07 \\
    SWBD-test & \textbf{15.02} & 15.18 & 15.22 & 15.48 & 15.51  & \textbf{11.93~(-20.57\%)} & 10.51  \\    
    TEDLIUM3-test & 4.22 & \textbf{3.95} & 4.26 & 4.51 & 4.69 & \textbf{3.40~(-19.43\%)} & 2.71  \\
    SPGI-val & \textbf{4.24} & 4.42 & 4.84 & 4.93 & 5.62 & \textbf{3.33~(-21.46\%)} & 2.15  \\
    \bottomrule
  \end{tabular}
  }
  \vspace{-5pt}
\end{table}

Table~\ref{tab:sentence_level} presents the results of sentence-level rescoring over the top-5 transcription hypotheses generated by Whisper large-v3. Across all datasets, sentence-level reranking based on the READ metric consistently outperforms the original top-1 results, demonstrating READ’s ability to effectively exploit the latent potential within the $N$-best list.

\subsubsection{Segment-Level Combination}
\begin{table}[h]
\vspace{-5pt}
  \caption{System Combination performances. ``R.”, ``Sen.”, ``Seg.”, and ``R+.” denote the ROVER baseline, sentence-level combination, segment-level combination, and ROVER integrated with segment-level combination, respectively.}
  \vspace{-0.1in}
  \setlength{\tabcolsep}{2.4pt}
  \label{tab:segment_level_and_rover}
  \centering
  \scriptsize
  \resizebox{\linewidth}{!}{
  \begin{tabular}{ccccccccc}
    \toprule
    \multirow{3}{*}{\textbf{Dataset}} & \multicolumn{4}{c}{\textbf{Candidates}} 
        & \multirow{3}{*}{\textbf{R.}} 
        & \multicolumn{3}{c}{\textbf{Ours}}  \\
        \cmidrule(lr){2-5}\cmidrule(lr){7-9}
        & 
    \makecell[c]{Whisper\\large-v3} & \makecell[c]{Whisper\\medium} &
     \makecell[c]{NeMo} & \makecell[c]{Qwen2.5-\\Omni} &  & Sen. & Seg. & R+. \\

    \midrule
    LS-clean & 2.20 & 2.79 & \textbf{1.67} & 1.74 & 1.51 & 1.67 & 1.66 & \textbf{1.49}\\
    LS-other & 4.16 & 7.52 & 3.65 & \textbf{3.45} & 3.18 & 3.39 & 3.35 & \textbf{2.97} \\
    VCTK-noisy & 8.87 & 18.33 & 2.85 & \textbf{2.47} & 1.84 & 2.36 & 2.18 & \textbf{1.52} \\
    ASRU-test & 10.35 & 11.98 & 21.70 & \textbf{8.00} & 9.04 & 7.60 & \textbf{7.27} & 7.60 \\ 
    TALCS-test & 16.77 & 20.74 & 44.47 & \textbf{9.21} & 20.84 & 9.61 & \textbf{8.22} & 16.55\\ 
    SWBD-test & 12.57 & 14.53 & \textbf{5.16} & 12.78 & 10.36 & \textbf{5.23} & 5.53 & 8.65 \\ 
    TEDLIUM3-test & \textbf{3.37} & 6.69 & 4.21 & 3.91 & 3.85 & \textbf{3.27} & 3.34 & 3.71\\
    SPGI-val & 3.60 & 7.25 & 4.40 & \textbf{2.76} & 3.15 & 3.14 & 3.01 & \textbf{2.80} \\
    \bottomrule
  \end{tabular}
  }
  \vspace{-5pt}
\end{table}

Table~\ref{tab:segment_level_and_rover} reports system combination performance across four candidate systems. On most datasets, segment-level combination method effectively leverages the locality of READ, achieving better performance through finer-grained merging.

On some datasets, however, segment-level combination does not significantly outperform sentence-level selection. This is because the disputed intervals are overly long, causing the segment-level combination to degenerate into sentence-level selection. A finer-grained scheme remains to be explored.

\subsubsection{Integrating READ with ROVER}

Table~\ref{tab:segment_level_and_rover} also compares ROVER integrated with READ. Although its performance is sometimes constrained by ROVER and thus cannot consistently surpass all alternatives, it steadily outperforms the vanilla ROVER baseline, demonstrating the effectiveness of READ as an incremental evaluation signal.

\subsubsection{Robustness against Noisy Speech}

\begin{table}[h]
\vspace{-5pt}
  \caption{System Combination performances on noisy datasets. ``Ours'' refers to direct segment-level combination without ROVER, since ROVER can be easily affected by several fragile systems.}
  \vspace{-0.1in}
  \setlength{\tabcolsep}{2.4pt}
  \label{tab:robustness_against_noisy_speech}
  \centering
  \scriptsize
  \resizebox{\linewidth}{!}{
    \begin{tabular}{c ccc ccc ccc}
        \toprule
        \multirow{2}{*}{\textbf{Dataset}} 
        & \multicolumn{3}{c}{\textbf{SNR=20dB}} 
        & \multicolumn{3}{c}{\textbf{SNR=10dB}} 
        & \multicolumn{3}{c}{\textbf{SNR=0dB}} \\
        
        \cmidrule(lr){2-4} \cmidrule(lr){5-7} \cmidrule(lr){8-10}
        
        & Best & ROVER & Ours
        & Best & ROVER & Ours
        & Best & ROVER & Ours \\
        
        \midrule
    LS-clean & 1.78 & \textbf{1.62} & 1.73 & 2.28 & \textbf{2.00} & 2.14 & 6.81 & 11.53 & \textbf{4.87}\\
    LS-other & 4.06 & \textbf{3.65} & 3.98 & 5.37 & \textbf{4.87} & 5.12 & 17.17 & 21.96 & \textbf{16.15}\\
    VCTK-noisy & \textbf{2.03} & 2.09 & 2.20 & 2.71 & 3.40 & \textbf{2.68} & 13.28 & 17.65 & \textbf{12.77}\\
    ASRU-test & 8.38 & 9.60 & \textbf{7.44} & 9.60 & 10.65 & \textbf{8.28} & 19.78 & 21.86 & \textbf{17.51}\\ 
    TALCS-test & 11.95 & 23.27 & \textbf{10.09} & 22.84 & 29.08 & \textbf{13.70} & 41.63 & 46.31 & \textbf{30.03}\\
    SWBD-test & \textbf{6.26} & 10.30 & 6.36 & 11.21 & 12.62 & \textbf{9.71} & 23.10 & 27.90 & \textbf{21.79}\\
    TEDLIUM3-test & 3.37 & 3.95 & \textbf{3.27} & \textbf{3.65} & 4.41 & 4.15 & 7.98 & 14.31 & \textbf{7.74}\\
    SPGI-val & 3.65 & 3.59 & \textbf{3.43} & 4.11 & 4.70 & \textbf{4.06} & 11.69 & 18.11 & \textbf{11.26}\\
        \bottomrule
    \end{tabular}
    }
  \vspace{-5pt}
\end{table}

As shown in Table~\ref{tab:robustness_against_noisy_speech}, as SNR decreases, the advantage of the proposed READ-based method becomes increasingly pronounced among the best single-system results and the ROVER outputs. This trend suggests that the effectiveness of READ-based evaluation is more evident under noisy conditions.

This is probably because in noisy conditions, ASR errors are more strongly related to acoustic modeling difficulty. In such cases, a powerful acoustic model is particularly important.

\section{Conclusion}

We propose READ, a hypothesis evaluation method based on the conditional likelihood modeled by a TTS system. READ is reference-free and requires only the original speech signal to assess hypotheses, indicating fine-grained acoustic discrepancy.

Without any additional training tailored to specific ASR models or datasets, our approach leverages the intrinsic knowledge of the TTS model to derive a READ metric that aligns well with specific recognition errors. This metric can be effectively applied to hypothesis refinement through $N$-best rescoring or system combination.

\section{Acknowledgments}
This work has been supported by the China NSFC Project (No. 92370206).

\section{Generative AI Use Disclosure}
Generative AI was utilized for manuscript editing and technical troubleshooting. The authors independently developed the research framework and experimental methodology. We take full responsibility for the content and consent to this submission.

\bibliographystyle{IEEEtran}
% \bibliography{main}

\begin{thebibliography}{10}
\providecommand{\url}[1]{#1}
\csname url@samestyle\endcsname
\providecommand{\newblock}{\relax}
\providecommand{\bibinfo}[2]{#2}
\providecommand{\BIBentrySTDinterwordspacing}{\spaceskip=0pt\relax}
\providecommand{\BIBentryALTinterwordstretchfactor}{4}
\providecommand{\BIBentryALTinterwordspacing}{\spaceskip=\fontdimen2\font plus
\BIBentryALTinterwordstretchfactor\fontdimen3\font minus \fontdimen4\font\relax}
\providecommand{\BIBforeignlanguage}[2]{{%
\expandafter\ifx\csname l@#1\endcsname\relax
\typeout{** WARNING: IEEEtran.bst: No hyphenation pattern has been}%
\typeout{** loaded for the language `#1'. Using the pattern for}%
\typeout{** the default language instead.}%
\else
\language=\csname l@#1\endcsname
\fi
#2}}
\providecommand{\BIBdecl}{\relax}
\BIBdecl

\bibitem{wessel2001confidence}
F.~Wessel, R.~Schluter, K.~Macherey, and H.~Ney, ``Confidence measures for large vocabulary continuous speech recognition,'' \emph{IEEE Transactions on speech and audio processing}, vol.~9, no.~3, pp. 288--298, 2001.

\bibitem{jiang2005confidence}
H.~Jiang, ``Confidence measures for speech recognition: A survey,'' \emph{Speech communication}, vol.~45, no.~4, pp. 455--470, 2005.

\bibitem{zhang2001word}
R.~Zhang and A.~I. Rudnicky, ``Word level confidence annotation using combinations of features,'' in \emph{Proc. Eurospeech 2001}, 2001, pp. 2105--2108.

\bibitem{guo2017calibration}
C.~Guo, G.~Pleiss, Y.~Sun, and K.~Q. Weinberger, ``On calibration of modern neural networks,'' in \emph{International conference on machine learning}.\hskip 1em plus 0.5em minus 0.4em\relax PMLR, 2017, pp. 1321--1330.

\bibitem{chelba2012large}
C.~Chelba, D.~Bikel, M.~Shugrina, P.~Nguyen, and S.~Kumar, ``Large scale language modeling in automatic speech recognition,'' \emph{arXiv preprint arXiv:1210.8440}, 2012.

\bibitem{mikolov2010recurrent}
T.~Mikolov, M.~Karafi{\'a}t, L.~Burget, J.~Cernock{\`y}, and S.~Khudanpur, ``Recurrent neural network based language model.'' in \emph{Interspeech}, vol.~2, no.~3.\hskip 1em plus 0.5em minus 0.4em\relax Makuhari, 2010, pp. 1045--1048.

\bibitem{Ali2018eWER}
A.~Ali and S.~Renals, ``Word error rate estimation for speech recognition: {e-WER},'' in \emph{Proceedings of the 56th Annual Meeting of the Association for Computational Linguistics (Volume 2: Short Papers)}, 2018, pp. 20--24.

\bibitem{rover}
J.~Fiscus, ``A post-processing system to yield reduced word error rates: Recognizer output voting error reduction (rover),'' in \emph{1997 IEEE Workshop on Automatic Speech Recognition and Understanding Proceedings}, 1997, pp. 347--354.

\bibitem{ma2023can}
R.~Ma, M.~Qian, P.~Manakul, M.~Gales, and K.~Knill, ``Can generative large language models perform asr error correction?'' \emph{arXiv preprint arXiv:2307.04172}, 2023.

\bibitem{chen2023hyporadise}
C.~Chen, Y.~Hu, C.-H.~H. Yang, S.~M. Siniscalchi, P.-Y. Chen, and E.-S. Chng, ``Hyporadise: An open baseline for generative speech recognition with large language models,'' \emph{Advances in Neural Information Processing Systems}, vol.~36, pp. 31\,665--31\,688, 2023.

\bibitem{tur2024progres}
A.~D. Tur, A.~Moumen, and M.~Ravanelli, ``Progres: Prompted generative rescoring on asr n-best,'' in \emph{2024 IEEE Spoken Language Technology Workshop (SLT)}.\hskip 1em plus 0.5em minus 0.4em\relax IEEE, 2024, pp. 600--607.

\bibitem{xu2024rejection}
H.~Xu, Z.~Zhu, S.~Zhang, D.~Ma, S.~Fan, L.~Chen, and K.~Yu, ``Rejection improves reliability: Training {LLM}s to refuse unknown questions using {RL} from knowledge feedback,'' in \emph{First Conference on Language Modeling}, 2024.

\bibitem{ali2020word}
A.~Ali and S.~Renals, ``{Word Error Rate Estimation Without ASR Output: e-WER2},'' in \emph{{Interspeech 2020}}, 2020, pp. 616--620.

\bibitem{park2025fast}
C.~Park, C.~Lu, M.~Chen, and T.~Hain, ``Fast word error rate estimation using self-supervised representations for speech and text,'' in \emph{ICASSP 2025-2025 IEEE International Conference on Acoustics, Speech and Signal Processing (ICASSP)}.\hskip 1em plus 0.5em minus 0.4em\relax IEEE, 2025, pp. 1--5.

\bibitem{waheed2025robust}
A.~Waheed, H.~Atwany, R.~Singh, and B.~Raj, ``On the robust approximation of asr metrics,'' in \emph{Findings of the Association for Computational Linguistics: ACL 2025}, 2025, pp. 23\,119--23\,146.

\bibitem{hu2024large}
Y.~Hu, C.~CHEN, C.-H.~H. Yang, R.~Li, C.~Zhang, P.-Y. Chen, and E.~Chng, ``Large language models are efficient learners of noise-robust speech recognition,'' in \emph{The Twelfth International Conference on Learning Representations}, 2024.

\bibitem{hu2024listen}
Y.~Hu, C.~Chen, C.~Qin, Q.~Zhu, E.~Chng, and R.~Li, ``Listen again and choose the right answer: A new paradigm for automatic speech recognition with large language models,'' in \emph{Findings of the Association for Computational Linguistics: ACL 2024}, 2024, pp. 666--679.

\bibitem{halle2003speech}
M.~Halle and K.~Stevens, ``Speech recognition: A model and a program for research,'' \emph{IRE transactions on information theory}, vol.~8, no.~2, pp. 155--159, 2003.

\bibitem{du2024cosyvoice2}
Z.~Du, Y.~Wang, Q.~Chen, X.~Shi, X.~Lv, T.~Zhao, Z.~Gao \emph{et~al.}, ``Cosyvoice 2: Scalable streaming speech synthesis with large language models,'' \emph{arXiv preprint arXiv:2412.10117}, 2024.

\bibitem{kim21e_interspeech}
S.~Kim, A.~Arora, D.~Le, C.-F. Yeh, C.~Fuegen, O.~Kalinli, and M.~L. Seltzer, ``{Semantic Distance: A New Metric for ASR Performance Analysis Towards Spoken Language Understanding},'' in \emph{{Interspeech 2021}}, 2021, pp. 1977--1981.

\bibitem{futami2021asr}
H.~Futami, H.~Inaguma, M.~Mimura, S.~Sakai, and T.~Kawahara, ``Asr rescoring and confidence estimation with electra,'' in \emph{2021 IEEE Automatic Speech Recognition and Understanding Workshop (ASRU)}.\hskip 1em plus 0.5em minus 0.4em\relax IEEE, 2021, pp. 380--387.

\bibitem{oneactua2021evaluation}
D.~Onea{\c{t}}{\u{a}}, A.~Caranica, A.~Stan, and H.~Cucu, ``An evaluation of word-level confidence estimation for end-to-end automatic speech recognition,'' in \emph{2021 IEEE Spoken Language Technology Workshop (SLT)}.\hskip 1em plus 0.5em minus 0.4em\relax IEEE, 2021, pp. 258--265.

\bibitem{wang2023neural}
S.~Chen, C.~Wang, Y.~Wu, Z.~Zhang, L.~Zhou, S.~Liu, Z.~Chen \emph{et~al.}, ``Neural codec language models are zero-shot text to speech synthesizers,'' \emph{IEEE Trans. ASLP}, vol.~33, pp. 705--718, 2025.

\bibitem{du2025cosyvoice}
Z.~Du, C.~Gao, Y.~Wang, F.~Yu, T.~Zhao, H.~Wang, X.~Lv, H.~Wang, C.~Ni, X.~Shi \emph{et~al.}, ``Cosyvoice 3: Towards in-the-wild speech generation via scaling-up and post-training,'' \emph{arXiv preprint arXiv:2505.17589}, 2025.

\bibitem{deng2025indextts}
W.~Deng, S.~Zhou, J.~Shu, J.~Wang, and L.~Wang, ``{IndexTTS}: An industrial-level controllable and efficient zero-shot text-to-speech system,'' \emph{arXiv preprint arXiv:2502.05512}, 2025.

\bibitem{moreno1998recursive}
P.~J. Moreno, C.~F. Joerg, J.-M. Van~Thong, and O.~Glickman, ``A recursive algorithm for the forced alignment of very long audio segments.'' in \emph{ICSLP}, vol.~98, 1998, pp. 2711--2714.

\bibitem{povey2011kaldi}
D.~Povey, A.~Ghoshal, G.~Boulianne, L.~Burget, O.~Glembek, N.~Goel, M.~Hannemann, P.~Motlicek, Y.~Qian, P.~Schwarz \emph{et~al.}, ``The kaldi speech recognition toolkit,'' in \emph{IEEE 2011 workshop on automatic speech recognition and understanding}.\hskip 1em plus 0.5em minus 0.4em\relax IEEE Signal Processing Society, 2011.

\bibitem{shi2026qwen3}
X.~Shi, X.~Wang, Z.~Guo, Y.~Wang, P.~Zhang, X.~Zhang, Z.~Guo, H.~Hao, Y.~Xi, B.~Yang \emph{et~al.}, ``Qwen3-{ASR} technical report,'' \emph{arXiv preprint arXiv:2601.21337}, 2026.

\bibitem{wang2024attention}
H.~Wang, C.~Du, Y.~Guo, S.~Wang, X.~Chen, and K.~Yu, ``Attention-constrained inference for robust decoder-only text-to-speech,'' in \emph{2024 IEEE SLT}.\hskip 1em plus 0.5em minus 0.4em\relax IEEE, 2024, pp. 630--637.

\bibitem{whisper}
A.~Radford, J.~W. Kim, T.~Xu, G.~Brockman, C.~Mcleavey, and I.~Sutskever, ``Robust speech recognition via large-scale weak supervision,'' in \emph{Proceedings of the 40th ICML}, vol. 202.\hskip 1em plus 0.5em minus 0.4em\relax PMLR, 23--29 Jul 2023, pp. 28\,492--28\,518.

\bibitem{nemofastconformer}
D.~Rekesh, N.~R. Koluguri, S.~Kriman, S.~Majumdar, V.~Noroozi \emph{et~al.}, ``Fast conformer with linearly scalable attention for efficient speech recognition,'' in \emph{2023 IEEE ASRU}.\hskip 1em plus 0.5em minus 0.4em\relax IEEE, 2023, pp. 1--8.

\bibitem{xu2025qwen25omnitechnicalreport}
J.~Xu, Z.~Guo, J.~He, H.~Hu, T.~He, S.~Bai, K.~Chen \emph{et~al.}, ``Qwen2.5-{Omni} technical report,'' \emph{arXiv preprint arXiv:2503.20215}, 2025.

\bibitem{librispeech}
V.~Panayotov, G.~Chen, D.~Povey, and S.~Khudanpur, ``Librispeech: An {ASR} corpus based on public domain audio books,'' in \emph{2015 IEEE ICASSP}, 2015, pp. 5206--5210.

\bibitem{spgispeech}
P.~K. O’Neill, V.~Lavrukhin, S.~Majumdar, V.~Noroozi, Y.~Zhang, O.~Kuchaiev \emph{et~al.}, ``{SPGISpeech: 5,000 Hours of Transcribed Financial Audio for Fully Formatted End-to-End Speech Recognition},'' in \emph{{Interspeech 2021}}, 2021, pp. 1434--1438.

\bibitem{godfrey1992switchboard}
J.~J. Godfrey, E.~C. Holliman, and J.~McDaniel, ``Switchboard: Telephone speech corpus for research and development,'' in \emph{IEEE ICASSP 1992}, vol.~1.\hskip 1em plus 0.5em minus 0.4em\relax IEEE, 1992, pp. 517--520.

\bibitem{tedlium3}
F.~Hernandez, V.~Nguyen, S.~Ghannay, N.~Tomashenko, and Y.~Esteve, ``{TED-LIUM} 3: Twice as much data and corpus repartition for experiments on speaker adaptation,'' in \emph{International conference on speech and computer}.\hskip 1em plus 0.5em minus 0.4em\relax Springer, 2018, pp. 198--208.

\bibitem{valentini2017vctknoisy}
\BIBentryALTinterwordspacing
C.~Valentini-Botinhao, ``Noisy speech database for training speech enhancement algorithms and {TTS} models, 2016,'' University of Edinburgh. School of Informatics. Centre for Speech Technology Research (CSTR), 2017, [sound]. [Online]. Available: \url{https://doi.org/10.7488/ds/2117}
\BIBentrySTDinterwordspacing

\bibitem{shi2020asru}
\BIBentryALTinterwordspacing
X.~Shi, Q.~Feng, and L.~Xie, ``The {ASRU} 2019 mandarin-english code-switching speech recognition challenge: Open datasets, tracks, methods and results,'' in \emph{Proceedings of the First Workshop on Speech Technologies for Code-switching in Multilingual Communities (WSTCSMC 2020)}, 2020, pp. 71--75. [Online]. Available: \url{http://festvox.org/cedar/WSTCSMC2020.pdf}
\BIBentrySTDinterwordspacing

\bibitem{talcs}
C.~Li, S.~Deng, Y.~Wang, G.~Wang, Y.~Gong, C.~Chen, and J.~Bai, ``{TALCS: An open-source Mandarin-English code-switching corpus and a speech recognition baseline},'' in \emph{{Interspeech 2022}}, 2022, pp. 1741--1745.

\bibitem{wham}
G.~Wichern, J.~Antognini, M.~Flynn, L.~R. Zhu, E.~McQuinn \emph{et~al.}, ``{WHAM!: Extending Speech Separation to Noisy Environments},'' in \emph{{Interspeech 2019}}, 2019, pp. 1368--1372.

\end{thebibliography}

\end{document}